\documentclass[manuscript,acmsmall,screen]{acmart}

\usepackage[utf8]{inputenc}
\usepackage{algorithmic}
\usepackage{textcomp}
\definecolor{Gray}{gray}{0.9}
\definecolor{kugray5}{RGB}{224,224,224}
\usepackage{acronym}
\usepackage{xspace}
\usepackage{multirow}
\usepackage{colortbl}
\usepackage{makecell}
\newcolumntype{g}{>{\columncolor{Gray}}c}
\usepackage{float}
\usepackage{pifont}
\newcommand{\cmark}{\ding{51}}%
\newcommand{\xmark}{\ding{55}}%
\usepackage[inline]{enumitem}
\usepackage[capitalise,nameinlink,noabbrev,english]{cleveref}

\newif\ifcommentcond
\commentcondfalse 

\newcounter{ms} 
\newcounter{mb} 
\newcounter{mat} 
\newcounter{me} 
\newcounter{hp} 
\newcounter{ant} 
\newcounter{fh} 
\newcounter{ab} 
\newcounter{wqy} 

\newif\ifparasumcond
\parasumcondfalse 
\newcommand{\parasum}[1]{\ifparasumcond \textcolor{teal}{#1}\fi}

\newcounter{thechalls}
\newcommand{\challenge}[1]{\refstepcounter{thechalls}%
  \paragraph{Challenge~\arabic{thechalls}: #1.}}
\makeatletter
\@ifpackageloaded{cleveref}{\crefname{thechalls}{Challenge}{Challenges}}{}
\makeatother
\newcounter{theresqs}
\newcommand{\resq}[1]{\refstepcounter{theresqs}%
  \paragraph{RQ~\arabic{theresqs}: #1.}}
\makeatletter
\@ifpackageloaded{cleveref}{\crefname{theresqs}{RQ}{RQs}}{}
\makeatother

\newcommand{\ignore}[1]{}

\newcommand{\etal}{et al.\xspace}

\newcommand{\systemname}{EmbedFuzz\xspace}

\newcommand{\evalRepetitions}{ten\xspace}
\newcommand{\evalTime}{24\xspace}
\newcommand{\fwbug}{eight}
\newcommand{\efbug}{seven}
\newcommand{\pimbug}{six}

\newcommand{\halfcirc}{\includegraphics[height=2ex]{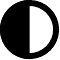}}
\newcommand{\fullcirc}{\includegraphics[height=2ex]{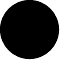}}
\newcommand{\emptycirc}{\includegraphics[height=2ex]{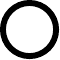}}

\newcommand{\arm}{Arm}

\acrodef{AAPCS}{\arm{} Architecture Procedure Call Standard}
\acrodef{ABI}{Application Binary Interface}
\acrodef{AFL}{American Fuzzy Lop}
\acrodef{APSR}{Application Program Status Register}
\acrodef{ASLR}{Address Space Layout Randomization}
\acrodef{CFG}{Control Flow Graph}
\acrodef{COTS}{Commercial-off-the-shelf}
\acrodef{CPU}{Central Processing Unit}
\acrodef{CPSR}{Current Program Status Register}
\acrodef{CWE}{Common Weakness Enumeration}
\acrodef{DEP}{Data Execution Prevention}
\acrodef{DMA}{Direct Memory Access}
\acrodef{DoS}{Denial-of-Service}
\acrodef{DSE}{Dynamic Symbolic Execution}
\acrodef{HAL}{Hardware Abstraction Layer}
\acrodef{HLM}{High-level Modeling}
\acrodef{ICS}{Industrial Control System}
\acrodef{I/O}{Input/Output}
\acrodef{IoT}{Internet of Things}
\acrodef{IR}{Intermediate Representation}
\acrodef{ISA}{Instruction Set Architecture}
\acrodef{ISR}{Interrupt Service Routine}
\acrodef{LSB}{Least Significant Bit}
\acrodef{MMIO}{Memory-mapped Input/Output}
\acrodef{MCU}{Microcontroller Unit}
\acrodef{MMU}{Memory Management Unit}
\acrodef{MPU}{Memory Protection Unit}
\acrodef{NVIC}{Nested Vectored Interrupt Controller}
\acrodef{OS}{Operating System}
\acrodef{OSS}{Open-Source Software}
\acrodef{PC}{Personal Computer}
\acrodef{PLC}{Programmable Logic Controller}
\acrodef{PUT}{Program Under Test}
\acrodef{RAID}{Redundant Array of Independent Disks}
\acrodef{RISC}{Reduced Instruction Set Computing}
\acrodef{RTOS}{Real-Time Operating System}
\acrodef{SBI}{Static Binary Instrumentation}
\acrodef{SDK}{Software Development Kit}
\acrodef{SMT}{Simultaneous Multithreading}
\acrodef{SoC}{System-on-Chip}
\acrodef{TDP}{Thermal Design Power}
\acrodef{UART}{Universal Asynchronous Receiver-Transmitter}

\begin{document}

\title{\systemname{}: High Speed Fuzzing through Transplantation}

\author{Florian Hofhammer}
\email{florian.hofhammer@epfl.ch}
\orcid{0009-0003-7661-3233}
\affiliation{%
  \institution{EPFL}
  \streetaddress{Station 15}
  \city{Lausanne}
  \postcode{1015}
  \country{Switzerland}
}

\author{Qinying Wang}
\orcid{0000-0002-0010-0592}
\email{qinying.wang@epfl.ch}
\affiliation{%
  \institution{EPFL}
  \streetaddress{Station 15}
  \city{Lausanne}
  \postcode{1015}
  \country{Switzerland}
}
\affiliation{%
  \institution{Zhejiang University}
  \city{Hangzhou}
  \country{China}
}

\author{Atri Bhattacharyya}
\email{atri.bhattacharyya@epfl.ch}
\affiliation{%
  \institution{EPFL}
  \streetaddress{Station 15}
  \city{Lausanne}
  \postcode{1015}
  \country{Switzerland}
}

\author{Majid Salehi}
\orcid{0000-0001-5781-7589}
\email{majid.salehi@nokia-bell-labs.com}
\affiliation{%
  \institution{EPFL}
  \streetaddress{Station 15}
  \city{Lausanne}
  \postcode{1015}
  \country{Switzerland}
}
\affiliation{%
  \institution{Nokia Bell Labs}
  \city{Antwerp}
  \country{Belgium}
}

\author{Bruno Crispo}
\orcid{0000-0002-1252-8465}
\email{bruno.crispo@unitn.it}
\affiliation{%
  \institution{University of Trento}
  \streetaddress{Via Sommarive 14}
  \city{Povo}
  \postcode{38123}
  \country{Italy}
}

\author{Manuel Egele}
\orcid{0000-0001-5038-2682}
\email{megele@bu.edu}
\affiliation{%
  \institution{Boston University}
  \streetaddress{Photonics Building, 8 St. Mary's Street, Room 337}
  \city{Boston}
  \state{Massachusetts}
  \country{USA}
}

\author{Mathias Payer}
\orcid{0000-0001-5054-7547}
\email{mathias.payer@nebelwelt.net}
\affiliation{%
  \institution{EPFL}
  \streetaddress{Station 15}
  \city{Lausanne}
  \postcode{1015}
  \country{Switzerland}
}

\author{Marcel Busch}
\email{marcel.busch@epfl.ch}
\affiliation{%
  \institution{EPFL}
  \streetaddress{Station 15}
  \city{Lausanne}
  \postcode{1015}
  \country{Switzerland}
}

\renewcommand{\shortauthors}{Hofhammer et al.}

\begin{abstract}
Dynamic analysis and especially fuzzing are challenging tasks for embedded
firmware running on modern low-end \acp{MCU} due to performance overheads
from instruction emulation, the difficulty of emulating the vast space of
available peripherals, and low availability of open-source embedded firmware.
Consequently, efficient security testing of \ac{MCU} firmware has proved to be a
resource- and engineering-heavy endeavor.

\systemname{} introduces an efficient end-to-end fuzzing framework for
\ac{MCU} firmware.
Our novel \textit{firmware transplantation} technique
converts binary \ac{MCU} firmware to a functionally equivalent and
fuzzing-enhanced version of the firmware which executes on a compatible
high-end device at native performance.
Besides the performance gains, our system enables advanced introspection
capabilities based on tooling for typical Linux user space processes, thus
simplifying analysis of crashes and bug triaging.
In our evaluation against state-of-the-art \ac{MCU} fuzzers, \systemname{}
exhibits up to eight-fold fuzzing throughput while consuming at most a
fourth of the energy thanks to its native execution.
\end{abstract}



\maketitle


\section{Introduction}
\label{sec:introduction}

\parasum{Embedded systems are ubiquitous and vulnerable. We want to
fuzz embedded systems because fuzzing has proven to effectively unveil memory
safety bugs.}
\ac{MCU}-based embedded systems present a large attack surface due to their
ubiquity, exposure to the internet, lack of \ac{MMU}-based
memory isolation, and development in memory-unsafe
languages like C/C++.
Vulnerabilities in \acp{MCU} used in devices like fitness trackers or smart
light bulbs~\cite{wegne2018fitbit,five2017withings,joshi2013philips} as well as
industrial applications~\cite{digikey2013mcus} put end users, industry, and society at large
at risk.
Fuzz testing~\cite{manes2021art,godefroid2020fuzzing} effectively uncovers
memory safety bugs in a wide range of software, from user space software over
\ac{OS} kernels to hypervisors~\cite{fioraldi2020afl,vyukov2022syzkaller,
bulekov2022morphuzz,liu2023videzzo,schumilo2021nyx}.
However, this promising testing technique faces several challenges when
targeting \ac{MCU} firmware.

\parasum{Challenge 1: Fuzzing embedded systems is slow $\Rightarrow$ emulation
tax}
\paragraph{Performance.}
Running \ac{MCU} fuzzing campaigns on server-grade hardware suffers from a
performance penalty due to incompatible \acp{ISA}.
For example, firmware targeting \ac{MCU} based on the \arm{}, AVR, or Xtensa
\acp{ISA} need to be translated dynamically to the native \ac{ISA} of
server-grade x86 processors.
Previous works~\cite{clements2020halucinator,muench2018avatar,zaddach2014avatar,
feng2020p2im,scharnowski2022fuzzware,zheng2019firmafl,zhou2021automatic}
leverage QEMU~\cite{bellard2005qemu} as a full-system emulator, consequently
paying a hefty \emph{emulation tax}.

\parasum{Challenge 2: Peripheral emulation is hard to get right
automatically/requires great engineering effort}
\paragraph{Peripheral Emulation.}
Embedded system firmware heavily relies on the \ac{I/O} capabilities of
peripherals attached to the corresponding \ac{MCU}.
Emulating peripherals in the re-hosting process either entails great
engineering effort~\cite{fasano2021sok} or requires sophisticated systems for
automatic detection and emulation of \ac{MMIO}
accesses~\cite{feng2020p2im,zhou2021automatic}, which may introduce
inaccuracies.
Forwarding  peripheral accesses to physical
hardware~\cite{zaddach2014avatar,muench2018avatar} hinders scalability due to
hardware dependence and incurs performance overheads stemming from
communication with the physical device.

\parasum{Challenge 3: Typically only firmware binaries $\Rightarrow$ no
source-based instrumentation, lacking tooling for static binary instrumentation}
\paragraph{Binary-only Firmware.}
Embedded system firmware for commercial platforms is seldom
available as \ac{OSS}.
Consequently, (re-)compiling firmware with fuzzing-enhancing instrumentation as
provided by off-the-shelf fuzzers such as AFL++~\cite{fioraldi2020afl} is only
possible for a small subset of firmware.
Low-overhead binary instrumentation is thus an important requirement for effective
fuzzing.
The dynamic binary instrumentation as provided by QEMU~\cite{bellard2005qemu},
for example in AFL++'s QEMU mode, comes with the emulation tax as described
above.
Static binary instrumentation, i.e., instrumenting the binary before its
execution, provides low runtime overhead and is ideal for introducing the
required instrumentation for fuzzing.

\parasum{Solution: EmbedFuzz, based on \emph{transplantation} and \ac{HLM}}
\paragraph{Our solution.}
We present \systemname{}, a high-performance, scalable, and accurate fuzzer for
embedded system firmware.
\systemname{} alleviates the emulation performance penalty by introducing and employing
\emph{transplantation}.
Transplantation
\begin{enumerate*}
  \item{%
    employs static binary rewriting on binary firmware images to make them
    compatible with an \ac{ISA} implemented in high-performance \acp{CPU}
    that are similar to the targeted embedded system, and
  }%
  \item{%
    executes the appropriately prepared bare-metal firmware as a user space
    process alongside a transplantation runtime on server-grade processors
    implementing such an \ac{ISA}.
  }
\end{enumerate*}
Through transplantation, \systemname{} leverages the native performance of
the underlying fuzzing hardware while at the same time scaling horizontally with
the typically high core count of such systems.
More specifically, our \systemname{} prototype transplants
\ac{MCU} firmware targeting \arm{} Cortex-M \acp{MCU} into Linux user space
processes executing on systems built around \arm{} Cortex-A processors.
Based on the observations made by HALucinator~\cite{clements2020halucinator},
\systemname{} identifies high-level interfaces to peripherals through binary
analysis, and builds on top of the principle of \ac{HLM} introduced by
HALucinator to emulate these interfaces.
Finally, \systemname{} enables the use of established tools for both fuzzing as
well as generic analysis and instrumentation purposes.
By running transplanted firmware as a native user space process, \systemname{}
can leverage \ac{OS} APIs such as \texttt{ptrace} used by existing tooling.

The key technique in our work is \emph{transplantation}, the process of moving
bare-metal firmware into a user space process.
This process is challenging due to architectural and environmental differences
between a bare-metal \ac{MCU} firmware and a common Linux user space process.
To address these challenges, \systemname{} leverages
\begin{enumerate*}[label=(\arabic*)]
  \item{modifications to the binary through static binary rewriting,}
  \item{execution in our user space runtime, and}
  \item{%
    similarity of the host \ac{ISA} and the \ac{ISA} of the targeted embedded
    system for native execution.
  }
\end{enumerate*}
In combination, \systemname{} through transplantation achieves
\begin{itemize}
  \item{%
    up to eight-fold increased fuzzing performance thanks to native execution
    of firmware binaries,
  }%
  \item{%
    horizontal scalability with core count thanks to \ac{MCU} hardware
    independence,
  }%
  \item{%
    reusability of existing tooling for common user space processes.
  }
\end{itemize}
\systemname{} is available as open-source software at
\url{https://github.com/HexHive/SURGEON}.


\section{Background}

\parasum{The structure of this sections is (1) embedded systems/\acp{MCU}, (2)
re-hosting, (3) \acp{HAL}, and (4) fuzzing.}
To provide the necessary background for \systemname{}' design and
implementation, we briefly highlight the main differences between
\ac{MCU}-based embedded systems and general-purpose \acp{CPU}.
We also introduce \acp{HAL} typically used in the development of embedded
system firmware, and showcase fuzz testing as a viable software testing
and bug finding technique.

\subsection{\acp{MCU} and General-purpose \acp{CPU}}

\parasum{MCUs do not have MMUs, have lower clock frequencies, optional
features. They're not suitable for general-purpose OSs.}
While \acp{MCU} share features such as arithmetic and logic capabilities with
other \acp{CPU}, they target different use cases than general-purpose \acp{CPU}
found in smartphones, tablets, or desktop computers.
Namely, \acp{MCU} are typically custom-tailored for low cost and energy
consumption.
Consequently, they run at several orders of magnitude lower clock speeds of
tens to a few hundred megahertz instead of multiple gigahertz.
Additionally, \acp{MCU} lack features such as \acp{MMU} altogether, or make
other features such as \ac{CPU} caches optional.
The latter is purely a performance optimization, but the former is a crucial
requirement for many general-purpose \acp{OS} to run on a \ac{CPU}.
For example, Windows, or macOS require an \ac{MMU} by design. While GNU/Linux
can also run in absence of an \ac{MMU}, the memory isolation through virtual
address spaces for processes requires an \ac{MMU} to be present.
Any of those \acp{OS} requires comparatively much more performance than
provided by an \ac{MCU} to provide a smooth user experience.
As a result, the main use case for \acp{MCU} is to run bare-metal software that
is either designed to run without any \ac{OS} abstractions at all or built
around embedded \ac{OS} flavors such as FreeRTOS~\cite{amazonfreertos} or
\arm{}~Mbed~OS~\cite{arm2022mbed}.

\parasum{\arm{} Cortex-M and Cortex-A are instances of MCUs and high-end CPUs
that
can be found in a plethora of devices.}
As an example, \arm{}'s Cortex-M processor family represents a group of 32-bit
\ac{RISC} \acp{MCU}.
Chips from that family can be found in industrial applications (e.g.,
\acp{PLC}~\cite{harris2016grbl,digikey2013mcus} and robots), consumer devices
(e.g., smart scales~\cite{five2017withings} and activity
trackers~\cite{wegne2018fitbit}), and the \ac{IoT} (e.g., smart light
bulbs~\cite{joshi2013philips} and thermostats).
On the other end of the spectrum, the \arm{} Cortex-A family of 64-bit
general-purpose \acp{CPU} powers
smartphones~\cite{qualcomm2022smartphone} as
well as high-end laptops, desktop computers, and
server platforms~\cite{apple2022mac,ampere2022ampere}.

\subsection{\acf{HAL}}

\parasum{HAL libraries provide a common interface across MCUs to firmware
developers and are therefore widely used.}
\acf{HAL} libraries are a common building block of \ac{MCU} firmware.
Such libraries provide abstractions for developers to interact with the target
system's hardware while hiding low-level implementation details and
functionality such as \ac{MMIO} address ranges for data exchange with
peripherals.
The provided interface is typically common across \acp{MCU} belonging to the
same family or even across families sold by the same vendor.
A single
\ac{HAL} library can thus cover a large number of distinct \acp{MCU}.
Consequently, firmware developed with \ac{HAL} libraries are less tightly tied
to a certain instance of the physical hardware.
For example, STMicroelectronics provides the STM32Cube \ac{SDK} that includes
\ac{HAL} libraries covering all 17 \ac{MCU} families comprising over 1,200
different configurations of their \arm{} Cortex-M based
microcontrollers~\cite{stmicroelectronics2022stm32cube,
stmicroelectronics2022stm32}.

\subsection{Re-hosting}

\parasum{Define re-hosting and mention why it is important. Allows the reader
to better understand what we're actually trying to do and why.}
\emph{Re-hosting}~\cite{fasano2021sok} is the
process of building a virtual environment for a given firmware image that
sufficiently models the firmware's hardware dependencies.
The goal of this process is to design this virtual environment in a way that
analyses conducted on the re-hosted firmware in this environment are
representative of executions on the real hardware.
Previous re-hosting approaches mainly focus on the difficulty of emulating
peripherals~\cite{feng2020p2im,clements2020halucinator,scharnowski2022fuzzware,
zhou2021automatic} but leverage off-the-shelf emulators such as
QEMU~\cite{bellard2005qemu} or Unicorn~\cite{quynh2015unicorn} for \ac{ISA}
emulation, thereby incurring emulation overhead.

Modeling peripherals in a re-hosting environment typically follows one of two
main angles.
Either accesses to memory-mapped peripheral registers in the \ac{MMIO} address
range are handled directly or \acf{HLM} is employed.
\ac{HLM} intercepts control flow in the firmware before the
access to the actual \ac{MMIO} region happens.
This interception occurs at a pre-defined higher level, e.g., at the level of
the \ac{HAL} API.
\ac{HLM}, as introduced by HALucinator~\cite{clements2020halucinator} and
employed in other systems~\cite{seidel2023forming}, allows peripheral modeling
with the semantics provided by the \ac{HAL}.
Instead of requiring to accurately mimick complex real hardware's behavior,
only the \ac{HAL} API's behavior with known semantics needs to be replicated.
A firmware's application logic calling into \ac{HAL} libraries is fully
oblivious to the logic executing below the \ac{HAL}.
Consequently, both \ac{HLM} and \ac{MMIO}-based peripheral modeling in this
case do not affect the semantics of the firmware's application logic.

\subsection{Grey-box Fuzzing}

\parasum{%
  Fuzzing is a test technique that repeatedly executes a software with
  different inputs, hoping to trigger new behavior.
  Grey-box fuzzing collects coverage to guide generation of new inputs towards
  interesting locations.
}
\emph{Fuzz testing} or \emph{fuzzing} is an automated testing technique
to find bugs in software by repeatedly executing the \acf{PUT} with new
inputs~\cite{sutton2007fuzzing,manes2021art,godefroid2020fuzzing}.
Ideally, those inputs trigger different behavior in the \ac{PUT} so that as
much of the \ac{PUT}'s \ac{CFG} as possible is covered and bug locations are
therefore visited.
If a bug is triggered at such a location, the fuzzer detects the bug either
through a crash of the \ac{PUT} or through timeouts if execution stalls.
In coverage-guided grey-box fuzzing, the \ac{PUT} is instrumented to report
back to the fuzzer when it reaches new locations.
This feedback allows the fuzzer to identify increased code coverage, which
in turn can be used for detecting and mutating interesting inputs to
exercise the newly reached part of the code.
As a result, this methodology has allowed fuzzers such as
\acs{AFL}++~\cite{fioraldi2020afl} or Syzkaller~\cite{vyukov2022syzkaller} to
unearth numerous bugs and vulnerabilities in both user space applications and
\ac{OS} kernels~\cite{heuse2022afl,vyukov2022syzbot}.


\section{Fuzzing \acs{MCU} Firmware Challenges}
\label{sec:challenges}

In this section, we detail the challenges of rehosting \ac{MCU} firmware that
we briefly mentioned in the introduction.
\Cref{tab:approach2chall} summarizes the comparison of different approaches.

\begin{table}[t]
  \begin{center}
    \caption{%
      Dynamically analyzing rehosted \ac{MCU} firmware is challenging. This
      table shows how existing approaches address these challenges.
      E - Emulation, N - Native, {\smaller\emptycirc{}} - badly supported,
      {\smaller\fullcirc{}} - well supported
    }

    \label{tab:approach2chall}
    \begin{tabular}{ c|cccc }
      Challenges &
      \rlap{\rotatebox[origin=l]{45}{Hardware-in-the-loop}}  &
      \rlap{\rotatebox[origin=l]{45}{Pure Emulation}}        &
      \rlap{\rotatebox[origin=l]{45}{High-level Modeling}}   &
      \rlap{\rotatebox[origin=l]{45}{Transplantation}}       \\
      \hline
      Phys. Addr. Space Repr.   & E/N & E & E & N   \\
      Architectural differences & E   & E & E & E/N \\
      Peripheral Handling       & N   & E & E & E   \\
      Insn Execution            & E   & E & E & N   \\
      Speed+Scalability         & \emptycirc{} & \halfcirc{} & \halfcirc{} & \fullcirc{}  \\
      Introspection             & \fullcirc{}  & \fullcirc{} & \fullcirc{} & \fullcirc{}  \\
      \hline
    \end{tabular}

  \end{center}
\end{table}

\challenge{Representation of the Physical Address Space}
\label{chall:address-space}
\parasum{Transplantation reproduces the physical address space required by
firmware in a virtual address space without requiring special configuration/an
additional translation step.}
\ac{MCU} firmware statically links all code and global data.
\acp{MCU} then map this firmware, its heap and stack as well as
\ac{MMIO}-mapped peripherals and control registers into a single flat address
space.
Those different memory regions for ROM, RAM or \ac{MMIO} are placed at fixed
locations in this physical address space.
Re-hosting solutions must account for the corresponding address space structure
to accurately model the firmware's expected execution environment.

Hardware-in-the-loop, pure emulation, and \ac{HLM} approaches all introduce a complex
emulation layer. This layer imposes a software-based address translation for every
access to the address space to compensate for any side effects. 
For example, accessing the \ac{MMIO} region requires peripheral-specific
handling of loads and stores, whereas accessing RAM requires only \ac{MPU} access
permission checks.
In summary, all accesses must
respect the original \ac{MCU}'s physical address space layout.

\challenge{\Ac{MCU} Architectural Features}
\label{chall:mcu-environment}
\parasum{MCUs exhibit system-level intricacies that need to be correctly
reproduced in a virtualized system, e.g., banked registers, exception entry and
exit routines, etc.}
\Acp{MCU} differ architecturally compared to a \ac{CPU}'s user
space mode that re-hosting solutions typically execute in.
Such differences include, e.g., execution modes, interrupt controllers
(\arm{}'s \ac{NVIC}), banked registers, or hardware-supported exception entry
and return routines not present in a Linux process.
Especially the latter presents a significant difference between \arm{} Cortex-M
\acp{MCU} and an unprivileged Cortex-A user space context.
Loading a value into the program counter register
always triggers a branch to that location in user space, while loading a magic value
(\texttt{0xFFFFFFF1} to \texttt{0xFFFFFFFD}) instead triggers a
hardware-implemented exception return routine on \arm{} Cortex-M.
Re-hosting must account for such features.

Previous hardware-in-the-loop~\cite{zaddach2014avatar,muench2018avatar},
\ac{HLM}~\cite{clements2020halucinator} or purely
emulation-based~\cite{feng2020p2im,scharnowski2022fuzzware} re-hosting
solutions commonly emulate the corresponding logic,
managing complex \ac{CPU} state solely in software.
Other approaches trying to execute \ac{MCU} firmware in user space ignore
this challenge altogether~\cite{li2021library,seidel2023forming}.

\challenge{Peripheral Handling}
\label{chall:peripherals}
\parasum{\ac{HLM} is easier to implement and performs better than
hardware-in-the-loop or manual implementation of peripheral emulators.}
\ac{MCU} firmware interacts with the outside world using peripherals.
A re-hosting environment must handle peripheral accesses with varying
degrees of accuracy, depending on the analyses to be conducted and how the
results of those analyses map to the real hardware.

Hardware-in-the-loop approaches are commonly the most accurate due to their
usage of "real" peripherals. Pure emulation provides high accuracy using the
peripheral's \ac{MMIO} interface. However, this approach suffers from
significant engineering efforts to correctly implement the vast amount of
available peripherals~\cite{fasano2021sok}.
\ac{HLM} leverages the \ac{HAL} layer of peripherals to reduce the engineering
effort for peripheral emulation.
However, as a consequence, this approach does not execute low-level peripheral
specific code beneath the \ac{HAL} layer.
The \ac{HAL} interface provides information about the semantics of
data being passed in and out of the firmware from and to peripherals, allowing
highly accurate modeling of those interfaces.
Peripheral modeling based on symbolic modeling or automated peripheral behavior
inference exposes a higher degree of automation but fails to guarantee the
same level of accuracy. 

\challenge{Correct Execution of Instructions}
\label{chall:instruction-semantics}
\parasum{Transplantation is the only approach that guarantees accurate (i.e.,
implemented (and hopefully tested) in hardware) execution of firmware code.}
The execution of instructions must be
functionally equivalent to the execution on the original \ac{MCU}.
While this generally does not pose an abundant problem, emulators and
hypervisors may introduce inaccuracy when modeling the behavior
of a physical \ac{CPU}~\cite{martignoni2012pathexplorationa,amit2015virtual,
ge2021hyperfuzzer}, impacting functional correctness of the re-hosting
environment.

\challenge{Speed and Scalability}
\label{chall:performance}
\parasum{Other approaches are slower than transplantation due to emulation.
Scalability is hurt by hardware-in-the-loop.}
Dynamic analysis techniques require speed and horizontal scalability.
For example, fuzzing benefits from fast and massively parallelized execution due
to more trials being executed in less time.

Hardware-in-the-loop approaches emulate the target's \ac{ISA} and forward
peripheral accesses to real hardware via debug probes.
They require dedicated \ac{MCU} hardware for each
running firmware instance.
Thus, these approaches lack scalability and suffer from a performance degradation due
to the forwarding of accesses to the physical device.
In contrast, pure emulation and \acf{HLM} approaches do not require dedicated
hardware and scale horizontally. Unfortunately, these approaches pay a
hefty emulation tax for running a foreign \ac{ISA} on the host's \ac{CPU}.

\challenge{Introspection}
\label{chall:introspection}
\parasum{Transplantation allows reusing user space tooling instead of requiring
additional software layers/hardware interfaces.}
Dynamic analysis requires flexible introspection capabilities.
Hardware-in-the-loop, pure emulation, and \ac{HLM} usually
require presence of a hardware debugging interface or modification of the
emulation layer, i.e., adding a debugger stub to enable attaching a debugger.
Therefore, these approaches result in additional engineering effort, a
restricted feature set in comparison to common interfaces for user space
software, communication overhead, or a combination of the previous.

\paragraph{Challenges Summary.}
We designed \systemname{} to address these challenges.
\systemname{} provides accurate modeling of \ac{MCU} behavior paired with high
execution speeds and simple horizontal scalability by combining accurate native
hardware behavior with the ease of handling minimal Linux user space processes.
As a result, our firmware transplantation approach, \systemname{}, outperforms
related approaches when fuzzing \ac{MCU} firmware.


\section{\systemname{} Design}
\label{sect:design}

\emph{Binary translation} captures the idea of translating from one execution
environment and \ac{ISA} to a fundamentally different \ac{ISA} and environment
combination. For instance, it enables the execution of an \arm{} Cortex-M
firmware on an x86-based machine leveraging an emulator. On the other hand,
\emph{direct execution} of a firmware on the intended \ac{MCU} provides an
analyst with firmware behavior exactly exhibiting the behavior of a real-world
deployment.

We design and build \systemname{} based upon the insight that the \arm{}
Cortex-M and Cortex-A \acp{ISA} overlap to a large extent.
Based on this observation, we introduce \emph{transplantation} as a middle
ground between binary translation and direct execution.
In transplantation, we leverage the semantic compatibility between the
\acp{ISA} to execute the majority of binary code in a Cortex-M firmware without
modification in a Linux user space process on a Cortex-A based system.
We address remaining syntactic and semantic incompatibilities through a
combination of static binary rewriting and a custom runtime, resulting in
minimal changes to the firmware binary.

Transplantation entails two major advantages.  First, \systemname{} achieves
high execution speeds thanks to native execution of firmware code.  Both binary
translation and direct execution of firmware suffer from low execution speeds.
The former requires an expensive architectural translation layer, while the
latter experiences the low clock speeds of \acp{MCU}.  Second, the significant
overlap between \ac{MCU} and \ac{CPU} \acp{ISA}, and careful handling of the
remaining differences provides \systemname{} with a means to replicate
firmware's behavior with high fidelity.

\begin{figure}
  \centering
  \includegraphics[width=0.6 \linewidth]{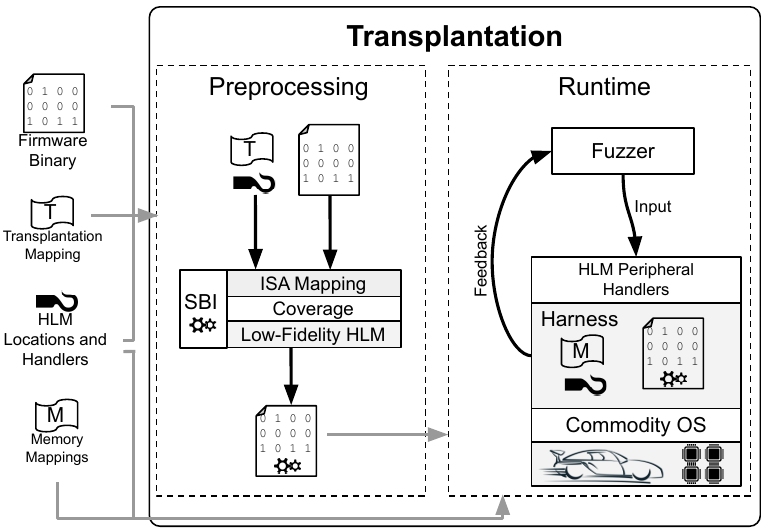}
  \caption{%
    The overview of \systemname{}'s approach to \ac{MCU} firmware fuzzing. Our
    contributions are shown in gray.%
  }
  \label{fig:embedfuzzdesign}
\end{figure}

\systemname{} leverages those insights for \emph{transplanting} embedded
firmware, creating fuzzable user space binaries in its \emph{transplantation
phase}.
\systemname{} handles peripheral accesses using \acf{HLM} in its \emph{runtime
phase}.
Together, transplantation and \ac{HLM} enable fast and accurate fuzzing while
tackling the challenges discussed in \cref{sec:challenges}.
We refer to \cref{fig:embedfuzzdesign} for an overview over \systemname{}'s
components.

Fuzzing embedded firmware with \systemname{} requires a \emph{transplantation
phase} that transforms the firmware into a fuzzable user space
binary through a sequence of four steps.
During this transplantation phase, \systemname{} modifies the binary to
\begin{enumerate*}[label=\emph{\roman*)})]
  \item{address semantic differences between the \acp{ISA} of
        Cortex-M and Cortex-A \acp{CPU} (\ac{ISA}
        mapping),}
  \item{insert coverage instrumentation (coverage),}
  \item{insert peripheral emulation code to directly replace simple \ac{HAL}
        functions (low-fidelity \ac{HLM}), and}
  \item{insert branches to the \ac{HAL} handlers in our runtime for complex
        \ac{HAL} functions (high-fidelity \ac{HLM}).}
\end{enumerate*}

After the transplantation phase, \systemname{}'s runtime ensures correct
execution of the prepared binary.
Our custom loader maps the binary with a virtual memory
layout matching the physical memory layout of the targeted \ac{MCU}, thereby
maintaining the validity of hardcoded addresses within the binary.
During analysis with a general-purpose coverage-guided fuzzer, \systemname{}
runs the firmware natively within a runtime which
\begin{enumerate*}[label=\emph{\roman*)})]
  \item{emulates architectural differences that cannot be addressed by
        static binary rewriting in the transplantation phase, and}
  \item{emulates the complex behavior of peripherals with \ac{HLM}.}
\end{enumerate*}
The runtime serves as the fuzzing harness, spawning a new process
representing the firmware's initial state and forwarding inputs from the fuzzer
for each iteration.
The runtime also monitors the process for anomalies (i.e., crashes and hangs)
in order to return feedback information (code coverage, termination status)
to the fuzzer.

In the following, we list the prerequisites for \systemname{}'s approach and
detail the design of the \emph{transplantation} and \emph{runtime} phases.

\subsection{System Model and Prerequisites}
\label{subsec:system-model}

\systemname{} targets firmware that lacks a classic \ac{OS} which would provide
process isolation and virtual address space abstractions.
%
Thus, \systemname{} assumes that the processes run on bare metal.
While the firmware binary must be available, the source code of the firmware
is not required.
%
Compared to emulator-based re-hosting techniques, transplantation has the
additional requirement that the high-end processor used for the fuzzing system
must natively support the simpler \ac{MCU}'s \ac{ISA}.
This requirement is satisfied by a wide variety of commodity platforms, with an
abundance of both embedded and high-performance processors based on \arm{},
PowerPC, or RISC-V.
Our prototype targets \ac{MCU} firmware developed for \arm{} Cortex-M.
This architecture is the most widely-deployed processor family for 32-bit
\acp{MCU}, and it is still actively used and deployed in new \ac{MCU} designs
\cite{almakhdhub2020rai,electronics2017reversal,salehi2019microguard,
kim2018securing}.
As detailed in \cref{subsec:transplantation}, \systemname{}'s design however is
not limited to this architecture.
Similar to HALucinator~\cite{clements2020halucinator} and
Safirefuzz~\cite{seidel2023forming}, we assume that \ac{HAL} libraries are
available to the analyst.
Cortex-M vendors provide open-source \ac{HAL} libraries for their chips, with
compilers and configurations.
Due to the benefits provided by these \ac{HAL} libraries, it is expected that most
developers utilize them~\cite{clements2020halucinator}.

\systemname{} discovers software
bugs and vulnerabilities in the firmware's code that can be triggered
by unexpected inputs provided to peripheral interfaces, e.g., SPI,
UART, or Ethernet. We leverage the full control over the inputs of such
exposed interfaces to inject fuzzing input into the firmware.

\subsection{Firmware Transplantation}
\label{subsec:transplantation}

\parasum{The reader now has seen the system model for \systemname{} and gotten
a quick introduction to its components. Now, we detail a bit more what
transplantation is and how this helps tackling the challenges in firmware
fuzzing.}
The principle of \emph{transplantation} is our key contribution to
enable high-performance execution of embedded system firmware binaries.
Transplantation requires ensuring correct execution of instructions as well as
mimicking the address space layout and architectural behavior that the
firmware expects.
Transplantation relies on the insight that the Thumb-2 instruction set
implemented by Cortex-M processors, which represent a dominant fraction of the
embedded device market, is similarly implemented in Cortex-A processors popular
in mobile and desktop \acp{CPU}.
Therefore, transplantation enables \systemname to execute embedded firmware
natively (i.e., without the cost of emulation) on Cortex-A processors
significantly faster than their original embedded system.
With desktop and server grade platforms incorporating up to 128
cores (e.g., for Ampere's Altra Max~\cite{ampere2022ampere}), transplantation
enables horizontal scaling of fuzzing campaigns.
Both those factors help \systemname{} address \cref{chall:performance} by
obviating performance overheads of running \ac{MCU} firmware either on-device
or in an emulator.
Native execution of firmware as a side effect also ensures correct execution of
instructions on real hardware (\cref{chall:instruction-semantics}).

\paragraph{Address space layout.}
Transplantation must preserve the address space layout of the
firmware since embedded code is typically statically linked with
hardcoded compile-time addresses.
Consequently, the firmware expects code, data, and other memory regions such as
\ac{MMIO} regions at certain locations in its physical address space
(\cref{chall:address-space}).
Our static binary rewriter accounts for this requirement by copying unmodified
instructions and data into the same location in the resulting binary.

\paragraph{\ac{MCU} environment modeling.}
As described in \cref{chall:mcu-environment}, a re-hosting solution must
sufficiently replicate the real device's behavior for accurate dynamic
analysis.
While the bulk of the \ac{MCU}'s instructions can be natively executed on
the fuzzing host, transplantation must account for the few instructions
 exhibiting different semantics on the two devices, including device-specific
effects such as programming interrupts, execution mode changes or
co-processor accesses.
\systemname accounts for these instructions by substituting them inline with
semantically equivalent instructions.
In the case where the inserted instructions would take up more space than the
original instruction in the binary did, \systemname{} follows a
trap-and-emulate approach and inserts traps to the runtime (see
\cref{subsec:runtime-env,subsubsec:implementation-transplantation}).

In addition to handling differing instruction semantics, transplantation models
\ac{MCU} hardware features not available on the transplantation host system
through additional architectural resources on high-performance \acp{CPU} to
address \cref{chall:mcu-environment}.
For instance, transplantation re-uses general purpose registers
not employed in the firmware for modeling \ac{MCU}-specific registers.
\systemname{} leverages this observation to implement features such as the
stack pointer register \texttt{sp} on Cortex-M being an alias to two banked
stack pointers used in the interrupt and standard mode, respectively.
We model this specific Cortex-M feature by mapping the stack pointer banks onto
vector registers only present on Cortex-A and rewriting instructions accessing
the corresponding registers accordingly.

\paragraph{Peripheral emulation.}
Transplantation must account for interactions with peripherals
in order to correctly model \ac{I/O} and interrupt behavior
(\cref{chall:peripherals}).
To this end, \systemname leverages HALucinator's high-level modeling
approach~\cite{clements2020halucinator}.
In the transplantation phase, \systemname{} inserts a branch to the
corresponding emulation code for each identified \ac{HAL} function.
As an optimization, simple \ac{HAL} functions such as peripheral initialization
functions are replaced inline with appropriate emulation code by our binary
rewriter (low-fidelity \ac{HAL}).

\paragraph{Fuzzing enhancements.}
Finally, \systemname{} optionally inserts fuzzing-specific instrumentation,
particularly edge coverage tracking, during static binary rewriting.
\systemname{} inserts trampolines to the instrumentation into basic blocks with
sufficient space for the corresponding branch instruction.
\systemname does not use the trap-and-emulate approach to track the remaining
basic blocks due to the high overhead of trapping.
With this design choice of opportunistic instrumentation, \systemname{}
trades performance for coverage accuracy in grey-box fuzzing campaigns.
As we empirically demonstrate in \cref{sec:evaluation}, this design choice is a
reasonable trade-off because the number of uninstrumented basic blocks in our
dataset is low and the fuzzer still manages to generate coverage outperforming
the competition.

\subsection{Runtime Environment}
\label{subsec:runtime-env}

\systemname{}'s runtime environment orchestrates the dynamic behavior required
to correctly fuzz the firmware.  First, the runtime sufficiently mimics the
\ac{MCU}'s behavior and handles emulation of complex instructions which trap in
the prepared binary to ensure correct execution of the firmware.  Second, the
runtime contains the \ac{HAL} handlers directly called by the firmware.  Third,
the runtime acts as a fuzzing harness for integration with general-purpose
fuzzers.

\paragraph{Address space layout.}
\systemname{}'s runtime ensures a correct address space representation
(\cref{chall:address-space}) by loading the firmware into a Linux process's
virtual address space according to the layout dictated by the \ac{MCU}'s
physical address space and the original firmware binary.

\paragraph{\ac{MCU} environment modeling.}
After loading the transplanted firmware binary, the runtime hands over control
to the natively executing firmware.
In order to correctly execute the firmware's code and model the \ac{MCU}'s
features (\cref{chall:mcu-environment,chall:instruction-semantics}),
\systemname{} employs multiple techniques.
First, the runtime handles traps issued during execution of the firmware in
case instructions with different semantics between \arm{} Cortex-M and Cortex-A
could not be substituted inline during the transplantation phase.
Second, the runtime faithfully emulates \ac{MCU}-specific behavior such as
saving and restoring execution context onto and from the stack or switching of banked
registers upon exception entry and return.
For example, loading magic values into the program counter on Cortex-M while in
an \ac{ISR} triggers a hardware-supported exception return routine restoring
context from the stack instead of an actual branch to the register value.
In a Cortex-A user space process, the \ac{CPU} will interpret those magic
values as addresses, try to execute code from that location, and raise a
segmentation fault.
By mapping branches to the corresponding software implementation of the
exception return routine at the addresses corresponding to the magic values,
\systemname{} avoids such false positive crashes and emulates the behavior
expected by the firmware.

Since embedded firmware generally process external interrupt-driven inputs, the
runtime must also correctly and reproducibly emulate interrupt behavior.
The reproducibility is essential for the developer to successfully triage
bugs.
Consider, for example, a timer which interrupts the firmware at regular
intervals in order to poll an attached \ac{UART} peripheral.
The runtime must periodically interrupt the firmware and call the
timer \ac{ISR} to emulate this behavior.
\systemname{} employs an interrupt scheduling approach similar to
P2IM~\cite{feng2020p2im} or HALucinator~\cite{clements2020halucinator},
delivering interrupts in a round-robin fashion at a fixed interval.
In contrast to the aforementioned systems, \systemname{} schedules interrupts
through a virtual clock based on instruction counting instead of basic block
counting.

\paragraph{Peripheral emulation.}
\systemname{}'s approach to peripheral emulation targets the higher-level
\ac{HAL} layer rather than the low-level \ac{I/O} instruction layer that
typically consists of \ac{MMIO} accesses, \ac{DMA} and interrupts.  By
leveraging the \ac{HLM} principle introduced by HALucinator, we consequently
allow replacing hardware peripherals with fully flexible software
implementations (\cref{chall:peripherals}).  In \systemname{}, the transplanted
binary calls into the runtime instead of the firmware's \ac{HAL}.  The runtime
integrates these calls with the corresponding emulation code.  Writing the
handlers for each \ac{HAL} function requires manual effort.  However, this
effort is acceptable as the cost amortizes across all firmware sharing the same
\ac{HAL} library.  Additionally, \systemname{} permits reusing existing handlers
from HALucinator~\cite{clements2020halucinator}.  These already existing
handlers provide a baseline to an analyst and can still be adapted according to
the use case.

\paragraph{Introspection.}
Since transplanted firmware runs inside a user space process, transplantation
facilitates the use of the plethora of dynamic analysis tools developed for user
space processes without restrictions or additional engineering requirements
(\cref{chall:introspection}).  For example, unmodified fuzzers such as
AFL(++)~\cite{zalewski2017american, fioraldi2020afl} or memory leak checkers
such as Valgrind~\cite{nethercote2007valgrind} can be directly applied to a
transplanted firmware.  Often, similar tools for \acp{MCU} and emulation-based
systems do not exist, or require additional steps or setup overhead.  Resetting
an \ac{MCU}'s state between fuzzing iterations is slow and requires hardware
debuggers with JTAG support, whereas the corresponding transplanted user space
process can be replicated with a fork system call.  Horizontal scaling for
transplanted firmware amounts to spawning additional processes on multi-core
systems (\cref{chall:performance}).  In contrast, similar scaling with \acp{MCU}
requires procurement of additional hardware and its orchestration.


\section{Implementation}
\label{sec:implementation}

Our \systemname{} prototype implements firmware transplantation as described in
the previous section for fuzzing firmware based on \arm{}'s Thumb-2 ISA.
Here, we describe key details of the implementation of our \systemname{}
prototype which we evaluate in \cref{sec:evaluation}.

\paragraph{Code Additions.}
\systemname{} maps added runtime code (for instruction emulation, peripheral
handling or fuzzing instrumentation) into an address range reserved for vendor
code as per the \arm{} Manual~\cite{arm2021armv7m}.
The \texttt{Vendor\_SYS} region, covering the address range
\texttt{0xE0100000-0xFFFFFFFF}, is unused in each of our analyzed firmware.
Even if a vendor makes use of parts of this memory range, it provides ample
space for storing any required code.

\paragraph{Transplantation.}
\label{subsubsec:implementation-transplantation}
For \systemname{}, we designed a custom static binary analysis and
instrumentation framework to transplant firmware and generate the fuzzing
binary.
\systemname{} transplants \ac{MCU} firmware, replicating non-code sections
and generating equivalent code for code sections using the methodology
described previously: copying the majority of instructions, and either
rewriting or emulating special instructions.
Through this approach, \systemname{} ensures semantically correct firmware
execution on an application processor, which is crucial for the accuracy of a
transplantation-based re-hosting system.

We determined the subset of special instructions by manually comparing the
Cortex-M and Cortex-A Thumb-2 instruction sets.
We first identified syntactically different instructions by forcefully
disassembling any possible two- and four-byte binary values that could
constitute an instruction and then comparing their mnemonics.
The results confirmed our assumption that the set of Thumb-2
instructions supported on Cortex-A \acp{CPU} is a superset of the one
encountered in our Cortex-M firmware dataset.
We then classified the instructions according to their behavior.
While it is safe to assume that the class of basic integer arithmetic
instructions such as \texttt{add} or \texttt{sub} behaves the same way on a
Cortex-M and a Cortex-A chip, other classes such as interrupt-generating
instructions (e.g., \texttt{svc}) require manual inspection.
This one-time effort does not need to be repeated for further use of our
system.

\systemname{} identifies three main categories of special instructions.
The first category consists of instructions that generate software interrupts.
For example, the execution of an instruction to change the privilege level
(e.g., \texttt{svc} (supervisor call)) might behave differently on a high-end
device.
The second category contains instructions that change the state of the \ac{CPU}
and usually require the current execution mode to be privileged.
An example for this category is enabling or disabling interrupt reception for
the current core (e.g., \texttt{cpsi\{e,d\}} (change processor state)).
In the third category, we find instruction-based \ac{I/O} which grants access
to \ac{CPU}-external resources using specific instructions (e.g., \texttt{mcr}
(move to coprocessor from \arm{} register)).
Accesses to co-processors using explicit instructions instead of \ac{MMIO} fall
into this category.
We provide further details on instructions requiring special handling in
\cref{app:semantically-different-instructions}.

\systemname{} replaces special instructions with semantically equivalent
instructions or, if not possible due to space constraints at the
instrumentation location, with traps.
The emulation code and trap handlers reside in the \texttt{Vendor\_SYS} region.
For trapping, we use the \texttt{bkpt} instruction, which causes the \ac{OS} to
raise a \texttt{SIGTRAP} signal to the runtime.
The runtime in turn implements the corresponding signal handler.
We uniquely identify each instruction (opcode and operands) to be emulated with
the \texttt{bkpt} instruction's 8-bit immediate field, allowing the handler to
dispatch to the correct emulation routine by reading this immediate field.
For example, supervisor call instructions (\texttt{svc}) are translated to
\texttt{bkpt \#1} instructions.
Apart from specific instructions that cannot be replaced inline, we also
replace infinite empty loops (i.e., a branch to itself) with a \texttt{bkpt} to
fast-forward \systemname{}'s virtual clock without stalling execution in such
loops.
We observe that firmware uses this pattern of an infinite loop for purely
interrupt-based logic, i.e., the main control flow of the firmware depends on
interrupts triggered by timers and other peripherals.
By fast-forwarding our internal virtual clock, we immediately trigger the next
interrupt while upholding the firmware logic's perception of time passed.
This is in contrast to previous and concurrent
work~\cite{scharnowski2022fuzzware,seidel2023forming}, which do not detect and
handle such loops waiting for interrupts.

While the \texttt{bkpt} approach limits \systemname{} to trap up to 256
different instruction types due to the size limitation of the immediate
operand, we found this sufficient for our fuzzed firmware.
An alternative implementation, adding complexity, can leverage a lookup table
to determine the instruction replaced with a \texttt{bkpt} and differentiate
based on the actual instruction instead of our \texttt{bkpt} operand.
Thus, this limitation is a conscious simplifying implementation choice.

\paragraph{Interrupts and Exceptions.}
\systemname{} emulates asynchronous software, \ac{MCU} and peripheral-driven
interrupts by periodically calling the corresponding handler in firmware based
on the programmed configuration in the interrupt vector table.
Handlers for \ac{HAL} functions configuring interrupts set up timers emulated
in our runtime which deterministically and reproducibly fire interrupts based
on the number of instructions executed.
Of particular importance, \systemname{} emulates the configurable timer
interrupt (\texttt{SysTick}) oftentimes leveraged for preemptive task
scheduling in \acp{RTOS}.
This approach allows \systemname{} to have reproducible fuzzing campaigns.

Importantly, \systemname{} also implements the \arm{} Cortex-M exception
entry and return routines executed by hardware on a real \ac{MCU}.
\systemname{} stores and restores context information to/from the stack as the
hardware would~\cite{arm2021armv7m} and also emulates switching between the
banked stack registers \texttt{SP\_main} and \texttt{SP\_process}.
The latter can be achieved very efficiently by re-purposing the Cortex-A's
double-precision floating point registers from \texttt{d16} onward, which are
not available on Cortex-M \acp{MCU}~\cite{arm2021armv7m,arm2022arm}.
Consequently, we replace the \texttt{mrs/msr} (move from/to special register)
instructions accessing the banked stack pointers with a single \texttt{vmov}
instruction in the binary rewriter, and switch stack pointers before/after an
\ac{ISR} implemented via two \texttt{vmov}s for storing/restoring the
corresponding banked version of the stack pointer.
This feature is crucial for supporting \ac{RTOS}-based \ac{MCU} firmware, since
task switching in \acp{RTOS} is commonly implemented by setting the stack and
banked stack registers up in an interrupt routine in such a way that the
hardware-implemented exception return routine switches to the desired stack and
restores the corresponding context from the stack.

\paragraph{Fuzzing Enhancements.}
\systemname{} implements support for AFL++ by inserting trampolines to record
coverage with basic-block granularity.
Basic blocks are identified with Ghidra~\cite{ghidra2022ghidra}, and the
corresponding coverage instrumentation is mapped into the \texttt{Vendor\_SYS}
region.
For certain basic blocks, we need to replace multiple Thumb-2 instructions
(typically 2 bytes) to fit the required 4-byte branch instruction.
A minority of basic-blocks ($\sim 10\%$), such as one holding a single
\texttt{ret} instruction, are too small to be instrumented, and are ignored.
Note that, while reducing accuracy, ignoring some basic blocks does not
overestimate the number of basic blocks covered.
Consequently, coverage reported by our instrumentation is a lower bound on the
actual coverage achieved.

\paragraph{Peripheral Handling.}
\systemname{} builds on HALucinator's~\cite{clements2020halucinator}
\acf{HLM}-based approach to emulating peripherals.
We use any symbols exported by the firmware or fall back to
HALucinator's \texttt{LibMatch} to identify \ac{HAL} function calls
and replace them with calls into the runtime's handler functions.
Our runtime implements \ac{HAL} handlers as C functions following the publicly
documented function prototype for the corresponding \ac{HAL} function.
Thanks to the adherence to the function prototype and the standardized \ac{ABI}
in the \ac{AAPCS}~\cite{armlimited2022procedure}, our handlers can consequently
access \ac{HAL} functions' arguments and return values directly without the
need of context modifications in between the invocation of the \ac{HAL}
function in the firmware and the execution of the corresponding handler.

Alternatively, our runtime links against \texttt{libpython}, allowing rapid
prototyping of \ac{HAL} handlers in Python and, more importantly, permitting
reusing HALucinator's peripheral handlers directly as a baseline for further
implementation.
The embedded Python interpreter can be used to execute peripheral handlers
written in Python in-process, thus not requiring a costly context switch to
another instance of a Python interpreter.
In contrast, related approaches rely on Linux's \texttt{ptrace} functionality
or dispatches from QEMU into a different handler
process~\cite{clements2020halucinator} and therefore add expensive context
switches for each \ac{HAL} function call.

Smaller or simpler handlers, such as \texttt{HAL\_TIM\_OsConfig} used in
the STM32 \ac{HAL} library for \ac{MCU} clock configuration, can be
replaced inline in the firmware code region or simplified to an empty function.
In this specific example, hardware clock configuration is irrelevant for
\systemname{} since timers are emulated in the runtime instead of relying on
hardware clocks, providing an analyst with fine-grained control over the
firmware's perception of time.


\section{Evaluation}
\label{sec:evaluation}

We evaluate \systemname{} on real-world firmware and compare it
against other \ac{MCU}-based coverage-guided greybox fuzzers.
We also conduct experiments to show that our transplantation approach is
accurate, i.e., that \systemname{} does not cause crashes or other unexpected
behavior which are not originating from actual bugs in the firmware.
Our evaluation is guided by the following research questions
that directly support our earlier claims.

\resq{Throughput}
\label{rq:throughput}
What is \systemname{}'s throughput (i.e., executions/sec) in comparison with
other state-of-the-art coverage-guided greybox fuzzers for \ac{MCU} firmware?

\resq{Coverage over time}
\label{rq:coverage}
What code coverage does \systemname{} achieve during fuzzing campaigns in
comparison to the other systems we evaluate against?

\resq{Bug finding capabilities}
\label{rq:bug-finding}
Which bugs in our evaluation dataset are uncovered by \systemname{} and the
other evaluated systems?

\resq{Computational resources}
\label{rq:resources}
What is the impact of \systemname{}'s transplantation approach on computational
resources required for fuzzing campaigns?

\resq{Transplantation fidelity}
\label{rq:transplantation}
How does transplantation fidelity impact accurate high-performance dynamic
analysis of \ac{MCU} firmware?

\resq{Scalability}
\label{rq:scalability}
What is the amount of manual work required by an analyst to transplant a
firmware with \systemname{} and how does this affect scalability to other
firmware?

\subsection{Experimental Setup}
\label{subsec:eval-setup}

In our experiments, we use 10 real-world firmware samples presented in
P2IM~\cite{feng2020p2im}.
These are full-fledged \ac{MCU} firmware and power various embedded devices
such as \acp{PLC}, drones, and robots.
They contain all the common \ac{MCU} firmware components (e.g., interrupt
handlers
and system libraries) and collectively cover \arm{} Cortex-M3 and Cortex-M4
microcontrollers.
To evaluate \systemname{}'s effectiveness and efficiency, we compare it with
two state-of-the-art coverage-guided greybox fuzzers for \ac{MCU} firmware:
P2IM~\cite{feng2020p2im} and Fuzzware~\cite{scharnowski2022fuzzware}.
Unfortunately, we encountered multiple issues in the
$\mu$Emu~\cite{zhou2021automatic} codebase%
    \footnote{git commit
    \href{https://github.com/MCUSec/uEmu/tree/a191402f5704fc231a7590fad7bc118f4208616b}{a191402f5704fc231a7590fad7bc118f4208616b}}
that prevented us from successfully
running all the desired experiments. Based on several weeks of engineering
efforts, some crashes in the S2E symbolic execution framework leveraged by
$\mu$Emu could be resolved thanks to correspondence with the author, other
problems such as excessive memory usage causing out-of-memory situations still
persist.  Those problems are also documented in the project's public GitHub
repository and have been encountered independently by multiple users.  In
consequence, we do not consider $\mu$Emu in our evaluation.
We do not evaluate against HALucinator's fuzzing component, hal-fuzz, and
Safirefuzz~\cite{clements2020halucinator,seidel2023forming}.
These two systems lack features modeling \ac{MCU} behavior related to
interrupts that transplantation provides and that are required for accurate
re-hosting of our evaluation dataset.
Consequently, the two systems are not compatible with our evaluation dataset.
We detail this observation in a case study in
\cref{subsec:case-study-soldering-iron}.

All experiments for transplanted firmware are conducted on a SolidRun Honeycomb
built around NXP's Layerscape LX2160A processor featuring 16 \arm{} Cortex-A72
cores.
The system is equipped with 32GB of memory and running the \arm{} version of
Ubuntu 22.04.
The dependencies of state-of-the-art \ac{MCU} fuzzers (e.g., AFL-Unicorn)
require that these are evaluated on machines powered by CPUs that implement the
Intel x86-64 ISA.
Thus, we use machines equipped with a 16-core Intel Xeon Gold 5218 processor
(hyperthreading disabled) and 64GB of RAM for their experiments.
These machines also run Ubuntu 22.04.
Furthermore, due to the randomness inherent in fuzzing, we conduct
\evalRepetitions \evalTime-hour trials of each fuzzing campaign and report
average, minimum and maximum numbers where appropriate, following established
guidelines~\cite{klees2018evaluating}.
For a fair evaluation, we kept the fuzzing parameters identical across all the
campaigns.
Each fuzzing campaign was configured with the same timeout, compute resources
and bootstrapped with the same set of seeds.
We conducted all experiments using our set of minimal input seeds, minimizing
the risk of impacting coverage increases due to seeds crafted with
firmware-specific knowledge.
We did not reuse seeds from P2IM or Fuzzware.
Instead, our initial minimal seed consists of the three-byte string
\texttt{foo}.
Seeds from related work are either hand-crafted to reach complex firmware
functionality or large undirected seeds (e.g., configuration files for
non-related software).
Starting with non-minimal seeds hinders comparability of the fuzzing campaigns,
because the fuzzers process the inputs differently depending on their
peripheral modeling strategy.
The start seeds might therefore allow one fuzzer to already bypass a difficult
condition that the other fuzzers get stuck on due to the different parsing of
the seeds.

\subsection{Fuzzing Throughput}
\label{subsec:fuzzing-throughput}

\begin{figure*}
  \centering
  \includegraphics[width=\linewidth]{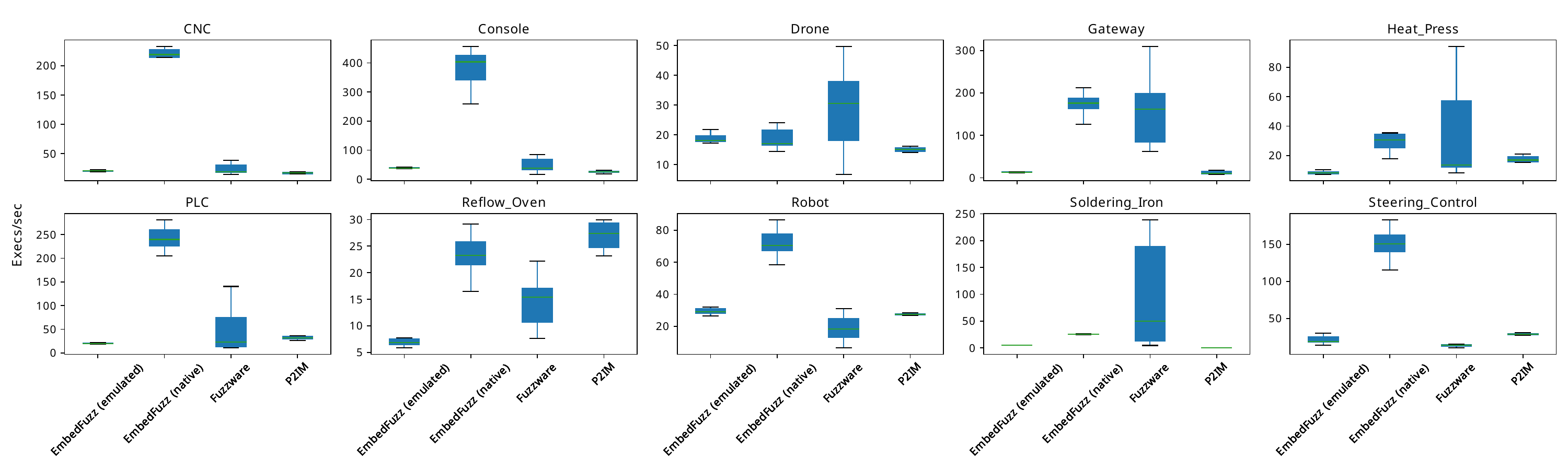}
  \caption{%
    Box plot of fuzzing executions per second (i.e., throughput) on real-world
    firmware binaries across \evalRepetitions \evalTime hour runs.
  }
  \label{fig:throughput}
\end{figure*}

We measured the throughput (i.e., the number of executions per second) of
\systemname{} compared to Fuzzware and P2IM.
Throughput is an important factor for fuzzing since exploring more inputs
directly correlates with a larger probability of discovering bugs.
\systemname{} improves fuzzing throughput by a factor of up to 8x in comparison
to its competitors, as shown in \cref{fig:throughput}.

As a notable exception, \systemname{} and Fuzzware both exhibit lower
executions per second on the Reflow Oven firmware than P2IM.
This observation can be explained through the excessive use of time
measurements in the firmware during which the firmware loops until an internal
counter incremented on each \texttt{SysTick} interrupt reaches a certain value.
We note that our virtual clock fast-forwarding design as described in
\cref{subsec:runtime-env} fast-forwards virtual time until the next
interrupt is triggered.
Our mechanism cannot account for internal counters managed by the firmware.
Our instruction counter based virtual clock in the configuration used for our
evaluation increments this counter slower than the implementations of Fuzzware
and P2IM.

\systemname{} shows similar throughput to Fuzzware on the Soldering Iron
firmware based on FreeRTOS~\cite{amazonfreertos}.
The complexity of FreeRTOS as further described in
\cref{subsec:case-study-soldering-iron} requires fallbacks to our
trap-and-emulate approach, incurring overheads through the context switches in
and out of the host \ac{OS} kernel during trap dispatching.

Based on our experiments, approximately 65\% of the handlers in MCU firmware
are trivial low-fidelity handlers that simply return a constant, such as the
ones used for hardware initialization functions.
As an example, configuring the \texttt{SysTick} timer in the STM32 \ac{HAL}
incurs calls to multiple functions for configuring system clocks, choosing
the clock to be used as a source for the \texttt{SysTick}, retrieving the
clock's frequency, and finally configuring the \texttt{SysTick} itself.
Only the last function in this call sequence is of interest for our \ac{HLM}
approach, since this is the only function that we concretely use for enabling the
emulated \texttt{SysTick} timer.
The previous functions therefore can without loss of functionality be replaced
with simple native handlers that return a value indicating successful execution
of the requested operation, speeding up execution.

Additionally, we compare the average number of executions per second of
\systemname{} running natively on \arm{} and on x86\_64 emulated with
\texttt{qemu-user}.
This experiment highlights the performance improvements through our
transplantation approach, resulting in an on average seven times higher fuzzing
throughput.

Based on the results of our experiments, we conclude that \systemname{}
exhibits higher throughput in executions/sec during fuzzing campaigns than
previous re-hosting systems, answering \cref{rq:throughput}.

\subsection{Coverage Over Time}
\label{subsec:cov-over-time}

\begin{figure*}
  \centering
  \includegraphics[width=\linewidth]{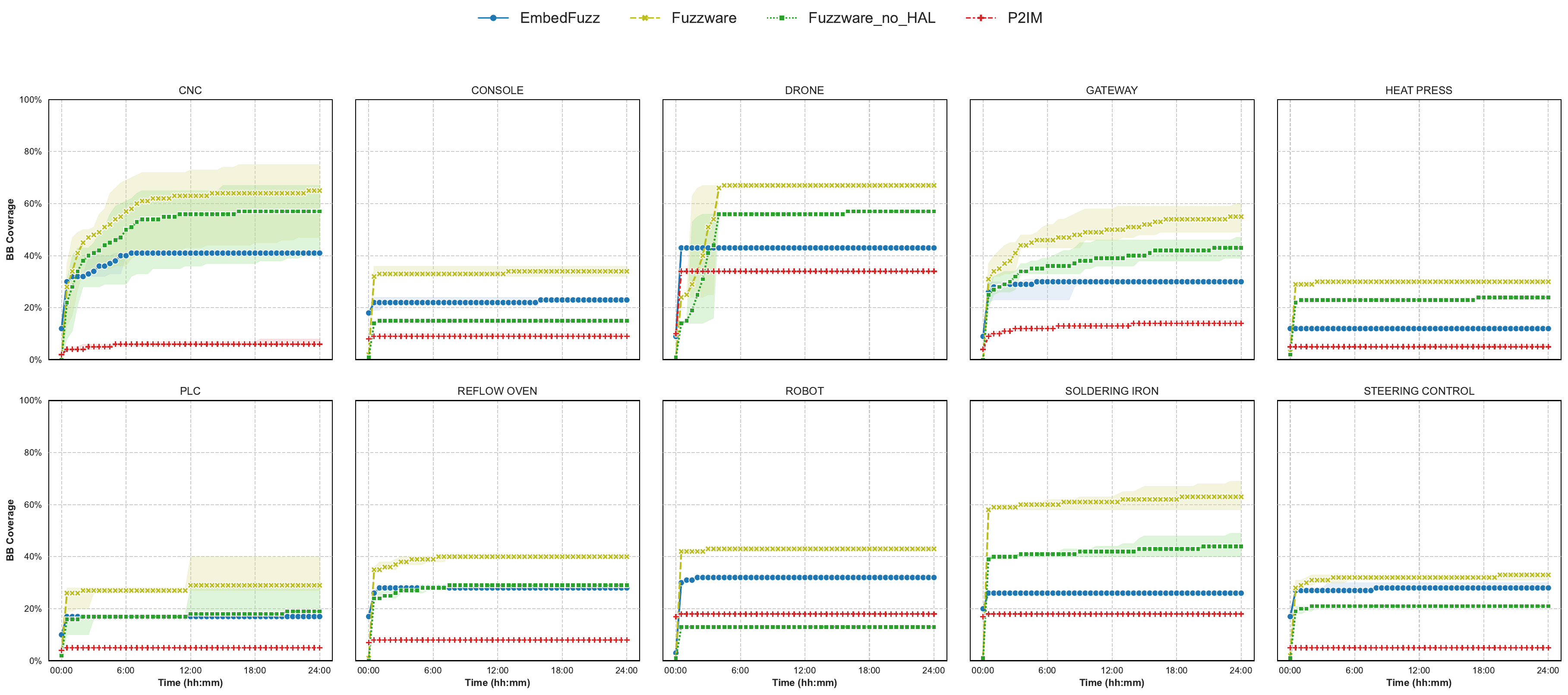}
  \caption{%
    Code coverage over time for real-world firmware binaries across
    \evalRepetitions \evalTime hour trials. The median percentage of uniquely
    discovered basic blocks, as well as the minima and maxima for each trial
    are illustrated.%
  }
  \label{fig:CovOverTime}
\end{figure*}

If a fuzzer covers more execution paths in the \acf{PUT} faster, it can earlier
probe the input space along these paths.
Thus, exploration is a prerequisite for finding bugs.
In our experiments, we measure the number of uniquely discovered basic blocks
by \systemname{} and compare against P2IM and Fuzzware over \evalRepetitions
\evalTime-hour runs.

Figure~\ref{fig:CovOverTime} summarizes our experiments on the number of
discovered basic blocks over time.
Given that Fuzzware and P2IM can reach many more basic blocks than
\systemname{} (e.g., basic blocks beneath the HAL layer), it is surprising to
see that \systemname{} outperforms P2IM in all ten cases.

For most of the firmware, Fuzzware covers
significantly more basic blocks than \systemname{} and P2IM.
Analysis of seed-generating traces shows that the majority of
these basic blocks are located in the \ac{HAL} libraries of the firmware.
However, as \systemname{} intercepts and redirects control at the entry of
\ac{HAL} APIs, the \ac{HAL} function implementations become unreachable code
for our approach.
We statically analyze the firmware binaries to determine basic blocks only
reachable from the \ac{HAL} functions \systemname{} intercepts.
We then evaluate the coverage attained by Fuzzware when subtracting \ac{HAL}
code as determined in our static analysis.
This approach provides a lower bound on the excluded basic blocks,
since targets of function pointers calculated at runtime cannot be reliably
detected.
As shown in \cref{fig:CovOverTime}, \systemname{} covers on par or more of the
firmware's code above the \ac{HAL} than Fuzzware in five out of ten cases.

For Console, Heat Press, and Robot, all fuzzers show signs of early
coverage stagnation.
A manual investigation revealed that there are conditional guards in the
firmware that are difficult to be satisfied with brute-force guessing.
Such guards are well-known to prevent fuzzers from reaching and covering deeper
parts of the code.
However, improving fuzzer capability to address challenges such as magic values
and checksums is outside the scope of this paper and has been the focus
of~\cite{cha2012unleashing,stephens2016driller,peng2018tfuzz}.

In summary, \systemname{} follows a high-level modeling approach and focuses on
the task logic above the \ac{HAL} of an embedded firmware.
Our coverage experiments show that \systemname{} can outperform P2IM's and
Fuzzware's coverage achieved during the fuzzing campaigns for our dataset,
positively answering \cref{rq:coverage}.

\subsection{Bug Finding and Debugging Capabilities}
\label{subsec:bug-finding}

\begin{table}
  \begin{center}
    \caption{%
      Bugs in our evaluation data set triggered by the three fuzzers.
      Bug ID refers to the identifier assigned by Fuzzware's experiments
      available at
      \url{https://github.com/fuzzware-fuzzer/fuzzware-experiments}.
      Abbreviations: \textbf{E}mbed\textbf{F}uzz; \textbf{F}uzz\textbf{W}are.
      Symbols: \cmark{} == bug triggered, \xmark{} == bug not reproducible,
      \textbf{?} == false positive bug%
    }
    \label{tab:bugscomparison}
    \small
    \begin{tabular}{cccccc}
      \toprule
      Bug ID & Target       & Description            & EF     & FW     & P2IM   \\
      \midrule
      11     & CNC          & Stack OOB write        & \cmark & \cmark &        \\
      12     & Gateway      & OOB write in HAL       & \cmark & \cmark & \cmark \\
      13     & Heat Press   & Buffer overflow        &        & \cmark & \cmark \\  
      14     & PLC          & Missing bounds check   & \cmark & \cmark & \cmark \\
      15     & PLC          & Missing bounds check   & \cmark & \cmark &        \\ 
      16     & PLC          & Missing bounds check   & \cmark & \cmark & \cmark \\  
      17     & PLC          & Missing bounds check   & \cmark & \cmark & \cmark \\
      18     & CNC          & CNC input validation   &        & \xmark &        \\  
      20     & Robot        & Initialization race    &        & \textbf{?} &    \\ 
      24     & Gateway      & Missing pointer check  & \cmark & \cmark & \cmark \\
      25     & PLC          & Missing initialization &        & \textbf{?} &    \\ 
      26     & Reflow Oven  & Missing pointer check  &        & \textbf{?} &    \\ 
      \midrule
      Total  &              &                        & 7      &  8     & 6      \\
      \bottomrule
    \end{tabular}
  \end{center}
\end{table}

We investigated the crashes generated by \systemname{}, Fuzzware, and P2IM.
Specifically, we first classified the crashing test cases generated
across experiments based on stack trace analysis.
Thereafter, we conducted a manual root-cause analysis for each group to
determine the type and severity of each bug.
The flexibility of \systemname{} in debugging helped us to readily perform
crash triaging.
For this use case, transplantation enabled us to utilize fully-fledged debugging
tools such as GDB. These tools are suitable for debugging native userspace applications without
going through debugging stubs in emulators or hardware debuggers.

In \evalTime hours of fuzzing of each firmware, \systemname{}, Fuzzware, and
P2IM reported \efbug{}, \fwbug{}, and \pimbug{} distinct bugs, respectively.
We provide details on the triggered bugs in \cref{tab:bugscomparison}.
Note that we focus on the subset of bugs reported by Fuzzware that are triggered
above the \ac{HAL} and that are hence reachable by \systemname{}.
However, the bug with ID 18 in Fuzzware's experiments was not reproducable.
The bugs with IDs 20, 25 and 26 are false positives only triggered due to
Fuzzware's interrupt modeling.
These bugs occur when an interrupt is triggered before it has been configured
in the firmware.
Fuzzware triggers available interrupts from the start of the execution onwards.
\systemname{} on the other hand only starts to trigger an interrupt once it has
been configured, which \systemname{} detects by hooking the corresponding
\ac{HAL} APIs.

Our analysis of the encountered bugs reveals oftentimes complex path
constraints to trigger a bug.
A symbolic execution-based fuzzer such as Fuzzware may overcome such
constraints more easily than other approaches thanks to its additional analyses.
We argue that this observation showcases the advantages and disadvantages of
different peripheral modeling approaches rather than transplantation.

In summary, \systemname{} was able to trigger \efbug{} bugs reported by Fuzzware,
and outperformed P2IM in its bug-finding capabilities, answering
\cref{rq:bug-finding}.
\systemname{}'s transplantation approach significantly simplified crash
triaging thanks to the availability of debugging tools for a normal user space
process instead of emulator-introduced debugging stubs.

\subsection{Power Efficiency}
\label{subsec:eval-power}

\parasum{Power efficiency is important.}
Power efficiency is crucial to modern datacenter design and must
be taken into account for large-scale fuzzing campaigns.
\parasum{The two CPUs under consideration are hard to compare because they're
so fundamentally different.}
NXP's LX2160A \ac{CPU} offers 16 cores with a
\ac{TDP} below 30W~\cite{halfhill2017lx2160a,nxp2020nxp}, while Intel's Xeon
Gold 5128 offers 16 cores with a \ac{TDP} of 125W~\cite{intelcorporation2017intel}.
Since the architectures and technical specifications of those \acp{CPU} as
detailed in \cref{tab:lx2160a-vs-xeon-gold-5128} are fundamentally different, a
power efficiency argument cannot be made on the \ac{TDP} alone.
For this reason, we measure the actual energy consumption of the full systems
running the fuzzing campaigns.
We encounter total system power draws of 47.0W and 197.1W on average for the
\arm{} and Intel x86\_64-based servers, respectively.
Our observations match the theoretical difference in \ac{TDP}, with an over
four-fold increase in power consumption for the Intel Xeon-based server in
comparison to NXP's LX2160A \ac{CPU}.

\parasum{When differentiating and taking all specs into consideration,
\systemname{} greatly increases power efficiency of fuzzing campaigns.}
Considering the LX2160A's lower operating frequency, the lower core count if
considering \ac{SMT}, and less cache memory available to the \ac{CPU}, NXP's
LX2160A \ac{CPU} at first glance seems to be less powerful than the Intel Xeon
Gold 5128 \ac{CPU} we use for the evaluation of \systemname{} against previous
work.
However, our numbers from \cref{subsec:fuzzing-throughput} show greatly
increased throughput when executing firmware natively with \systemname{}
instead of relying on emulation.
In our evaluation, \systemname{} therefore runs fuzzing campaigns with a 4-16
increased power efficiency depending on the analysis target in comparison to
emulation-based systems.
This observation holds true even when generously assuming that enabling
\ac{SMT} would double the processing power available to the Intel \acp{CPU}
without increasing power consumption.

\parasum{The increased power efficiency benefits data-center design and costs
of a fuzzing campaign.}
As a result, transplantation instead of emulation-based re-hosting has a
significant impact on the energy consumption of large-scale fuzzing campaigns,
while still increasing the campaigns' performance.
Lower energy consumption implies simpler and cheaper infrastructure due to
lower requirements, for example on cooling or power supplies.
The reduced cost for electrical power itself must also be taken into
consideration.
Consequently, \systemname{} significantly reduces the required computational
resources for fuzzing campaigns, answering \cref{rq:resources}.

\begin{table}
  \begin{center}
    \caption{
      Comparison of the NXP LX2160A~\cite{halfhill2017lx2160a,nxp2020nxp}
      and Intel Xeon Gold 5128~\cite{intelcorporation2017intel} \acp{CPU}.
      Power draw numbers refer to the full system's wall power draw, measured
      over 24 hours during fuzzing campaigns.
    }
    \label{tab:lx2160a-vs-xeon-gold-5128}
    \small
    \begin{tabular}{lrr}
      \toprule
                            & LX2160A     & Xeon Gold 5128            \\
      \midrule
      Architecture          & \arm{}v8-A  & x86\_64                   \\
      Cores                 & 16          & 16 (+16 with \acs{SMT})   \\
      Base frequency        & 2.2GHz      & 2.3GHz                    \\
      Turbo frequency       & 2.2GHz      & 3.9GHz                    \\
      Cache (L1 + L2 + L3)  & 18MB        & 22MB                      \\
      \acs{TDP}             & 30W         & 125W                      \\
      Power draw (max)      & 47.1W       & 200.0W                    \\
      Power draw (avg)      & 47.0W       & 197.1W                    \\
      Power draw (idle)     & 29.9W       & 84.0W                     \\
      \bottomrule
    \end{tabular}
  \end{center}
\end{table}

\subsection{C/Python Peripheral Handlers}
\label{subsec:c-vs-python-handlers}
%
\parasum{%
  Python handlers taken from HALucinator reduce implementation effort.
  C handlers provide significant speedups due to (1) the interpreted nature of
  Python and (2) the required in-process context switching between firmware and
  Python handlers.
}
\systemname{} supports peripheral handlers implemented both in C and Python.
C handlers in our implementation are required to follow the corresponding
\ac{HAL} function's prototype.
\arm{}'s \ac{ABI}, the \ac{AAPCS}, then ensures correct propagation of
arguments and return values as well as saving and restoring of callee-saved
registers.
Thanks to an embedded Python interpreter in the runtime, \systemname{} also
permits re-using HALucinator's peripheral handlers with minimal to no changes.
This feature allows for less engineering work to be conducted by a developer
and rapid prototyping thanks to the simplicity of the Python language.

However, natively compiled handlers provide significant performance advantages
over Python handlers, which is an important factor for high-performance dynamic
analysis.
In our experiments, fuzzing campaigns leveraging C handlers ran with up to 17x
improved executions/sec in comparison to their Python handler based
counterparts, with an average increase of 5x.
We attribute the slowdown incurred to the interpreted nature of the Python
programming language.
Additionally, we observed the Python interpreter to issue a multitude of system
calls for each imported Python module, both for locating and actually
interpreting the module.
These system calls, causing context switches into and out of the \ac{OS}
kernel, incur additional overhead.

Based on our experiments, we conclude that pre-existing Python handlers taken
from HALucinator provide an excellent base for quick prototyping, but that
peripheral handlers should be implemented in C to fully reap the benefits of
\systemname{}'s native execution.

\subsection{Case Study: The Soldering Iron Firmware}
\label{subsec:case-study-soldering-iron}
%
The Soldering Iron firmware from the P2IM dataset used for our evaluation is
based on IronOS~\cite{ironoscontributors2023ironos}.
This firmware is available for a variety of small, handheld soldering irons and
is built on top of FreeRTOS~\cite{amazonfreertos}.
The \ac{RTOS} runs multiple tasks, each with a dedicated responsibility such as
temperature sensing, power control, or handling UI tasks via builtin buttons
and screens.
The task scheduling functionality of an \ac{RTOS} such as FreeRTOS highlights
the importance of accurate system modeling in a re-hosting environment as
provided by \systemname{}'s transplantation approach.

First, the firmware executes an \texttt{svc} (supervisor call) instruction to
call into FreeRTOS's task scheduler once the system is initialized.
As described in \cref{subsubsec:implementation-transplantation}, \systemname{}
substitutes the \texttt{svc} instruction in the firmware binary with a trap to
our runtime in which we execute code mimicking the expected behavior.
Hal-fuzz, based on the Unicorn emulation engine~\cite{quynh2015unicorn}, does
not install hooks in the \ac{ISA} emulation engine for handling this
instruction, effectively ignoring its effects.
Safirefuzz also neglects to handle this instruction but causes side effects on
the host system due to native execution of the instruction.
On an \arm{}-based Linux system, a supervisor call is the equivalent to x86's
\texttt{syscall} instruction, calling into the kernel and executing unintended
system calls.

Second, once an interrupt is triggered on Cortex-M, the \ac{CPU} pushes
execution context onto the stack, executes the interrupt handler as defined in
the vector table, and finally pops execution context from the stack before
continuing execution of user code.
FreeRTOS leverages this hardware-supported context save and restore around
\acp{ISR} to implement task switches by setting the stack for each task up
accordingly.
Once the initial code triggers a software interrupt via the \texttt{svc}
instruction, the corresponding interrupt handler only needs to switch the stack
pointer to point to another task's stack.
Then, the \ac{ISR} simply returns, after which the \ac{CPU} restores the new
task's context from the stack.
\systemname{} supports this behavior in software, pushing and poping execution
context to and from the stack around the \ac{ISR} in a few assembly
instructions.
Hal-fuzz also implements these exception entry and return routines in software,
but needs to interact with the Unicorn \ac{ISA} emulator to modify its register
and memory state.
Safirefuzz lacks an implementation for replicating this hardware behavior.

Finally, FreeRTOS tasks leverage the PendSV interrupt to give up control and
let the scheduler switch to the next task.
While the interrupt triggered by an \texttt{svc} instruction is synchronous,
i.e., immediately triggered, the PendSV interrupt is asynchronous.
The firmware sets a flag in a system control register to mark the interrupt as
pending and it is triggered once no interrupts with a higher priority need to
be served anymore.
\systemname{} adds instrumentation to the trampolines used for fuzzing coverage
collection and virtual clock emulation, which checks whether this flag is set.
If it is, the same context save and restore routines as described above are
executed around the \ac{ISR} for the PendSV interrupt.
Both hal-fuzz and Safirefuzz are do not replicate this behavior of a real
\ac{MCU}.

While this case study focuses on a FreeRTOS-based firmware, other \acp{RTOS}
such as RIOT used in the Console firmware in our dataset follow the same model
for task switching.
In consequence, both hal-fuzz and Safirefuzz are incompatible with the P2IM
firmware data set due to shortcomings in modeling the \ac{MCU}'s behavior.

In response to \cref{rq:transplantation}, the fidelity of transplantation for
high-performance dynamic analysis of embedded firmare, we argue that
\systemname{}'s prototype is the only approach providing both high execution
speeds based on native execution and semantically correct execution through
transplantation.

\subsection{Scalability}
\label{subsec:manual-effort}
%
Transplantation generally does not require manual engineering efforts from an
analyst.
The required manual tasks for peripheral modeling and harnessing in our
\systemname{} prototype solely stem from HALucinator's \ac{HLM} approach.
In this approach, modeling a peripheral is an initial one-time manual effort
which amortizes with each additional firmware that uses a modeled peripheral.
In our data set, seven out of ten firmware leverage STMicroelectronics' STM32
\ac{HAL}, highlighting the possibility to reuse peripheral handlers across
firmware.
We implement peripheral handlers in C, following the original \ac{HAL}
functions' interfaces.
The implementation of the 49 handlers required to support the firmware in our
data set resulted in $\sim1,000$ lines of C code.

As described in \cref{sec:implementation}, \systemname{} also supports
prototyping of peripheral handlers in Python based on HALucinator's
interface.
In addition to our C handlers, we ported peripheral handlers published with
HALucinator to showcase \systemname{}'s versatility.
Our usage of in-process handlers instead of QEMU for \ac{HAL} function
interception and our custom interrupt-handling mechanism required
straightforward modifications to some HALucinator handlers which took one day
of engineering effort.

Implementing handlers in C or Python requires minimal effort in the order of a
few hours per \ac{HAL} from an analyst familiar with the \ac{HAL}.
We consequently conclude that the one-time effort required for peripheral
modeling allows an analyst to easily scale \systemname{} to other firmware,
answering \cref{rq:scalability}.


\section{Discussion}
\label{sec:discussion}

\parasum{Short intro to the section: discuss limitations, future directions}
Even though \systemname{} advances the state of the art based on its
contributions highlighted in the previous sections, there are some limitations
to the current implementation.
We now discuss these limitations, their effect on the general
applicability of our system, and how those limitations can be addressed in
future work.

\parasum{
  Limitations due to HAL-based interception: \\
  (1) need firmware using HAL, \\
  (2) need HAL source access (either for direct matching with symbols or usage
  with LibMatch)
}

\paragraph{High-level Modeling.}
\systemname{}'s reliance on \ac{HLM} based on the approach provided by
HALucinator~\cite{clements2020halucinator} comes with similar limitations to
those presented by Clements~\etal{}.
Notably, the firmware to be transplanted needs to be implemented using
\ac{HAL} libraries or embedded \acp{OS} providing a similar interface in order
for \systemname{} to be able to intercept peripheral accesses at this
interface. We expect this to be the case for real-world firmware.
Without this prerequisite, \systemname{} cannot handle peripheral
accesses triggered by the targeted firmware.

\systemname{} does not require source code access for the full \ac{MCU}
firmware.
However, access to the \ac{HAL} function source code is required to match
symbols in the firmware binary to the corresponding function to correctly
implement our \ac{HLM} handler as described in \cref{sec:implementation}.
If symbols are not available, we use HALucinator's LibMatch approach which
correlates compiled \ac{HAL} functions with the provided binary to find the
correct offsets.
Both approaches require access to the \ac{HAL} function source.

Even though we encounter those limitations, we argue that handling peripheral
accesses on the \acl{HAL} reduces analysis effort due to the reusability of
peripheral models across firmware if they rely on the same \ac{HAL} library.
With the \ac{HAL} libraries being provided openly by \ac{MCU} vendors to
developers, reuse of \ac{HAL} libraries is widespread and \systemname{}
therefore greatly benefits from the reusability of its peripheral models.

\parasum{
  Limitations of transplantation: \\
  (1) need high-performance system with compatible ISA $\Rightarrow$ works well
  for \arm{}, maybe also for PowerPC, but becomes more difficult for AVR,
  Xtensa,
  etc., \\
  (2) high-performance CPU must have superset of MCU features, otherwise we
  need to fallback to instrumentation (as seen for certain instructions) or
  emulation (in case of completely missing features), \\
  (3) don't have the possibility of using interrupts as done on native
  firmware but need to fall back to a mixture of inline code and runtime traps
  and corresponding handlers
}

\paragraph{Transplantation and Native Code Execution.}
One of the features of \systemname{}'s transplantation approach is the
performance improvement through native code execution instead of emulation.
As outlined before, this requires a \ac{CPU} supporting the target firmware's
\ac{ISA}.
In our implementation, we run code targeting \arm{} Cortex-M
\acp{MCU} on \arm{} Cortex-A \acp{CPU}.
\systemname{}'s implementation, by design, is also able to execute code
targeting \arm{} Cortex-R based systems as long as the firmware does not
leverage an \ac{MMU}. In this case, the firmware's physical address space can be
modeled in our user space process's virtual address space without additional
address translation costs, as described in \cref{subsec:runtime-env}.
While this approach can certainly be extended to other architectures such as
PowerPC featured in \acp{MCU} (e.g., NXP's MPC5xxx
\acp{MCU}~\cite{nxp2022mpc5xxx}) as well as in high-performance server
\acp{CPU} (e.g., IBM's Power10 \acp{CPU}~\cite{ibm2022ibma}), it is not
applicable to all \ac{MCU} architectures.
To the best of our knowledge, there are no high-performance \acp{CPU} on the
market supporting the AVR or XTensa \acp{ISA} which excludes \ac{MCU} firmware
based on these architectures from the list of possible transplantation targets.

Transplantation is successful because we leverage a high-performance host
\ac{CPU} feature superset of the low-power \ac{MCU} features.
If a firmware makes extensive use of features not available in \arm{} Cortex-A
\acp{CPU} such as a \ac{MPU}, \systemname{} currently does not support
transplantation of such firmware. Our prototype can be extended to
emulate such behavior by inducing significant performance costs.

Another limitation is the handling of interrupts triggered by peripherals.
As described in \cref{subsec:runtime-env}, we simulate interrupts occurring at
regular intervals.
We relate the time elapsed to the number of executed instructions.
\systemname{} falls back to a runtime trap wherever this is not
possible, e.g., for software interrupts triggered via an \texttt{svc}
instruction.
Given peripheral interception, this may not accurately match the time executed
in real firmware.
This is an issue of re-hosting fidelity~\cite{wright2021challenges}.
As a result, \systemname{}'s interrupt simulation is considered a best-effort
approach that cannot simulate a real interrupt-based system's behavior with
100\% accuracy.
We nevertheless contend that our strategy is more than sufficient for our
needs.


\section{Related Work}
\label{related_work}

With the wide adoption of fuzzing,  a large body of related work has
contributed to applying fuzzing to embedded systems to enhance their
security.

We analyze existing work targeting embedded system fuzzing along several
dimensions:
\begin{enumerate*}[label=(\arabic*)]
  \item code execution,
  \item peripheral modeling approach,
  \item supported \ac{I/O} methods,
  \item dependency on hardware,
  \item fault detection mechanism, and
  \item dependency on source code.
\end{enumerate*}
Table~\ref{tab:compare} summarizes a selection of related work and tools,
representing all the different categories of approaches laid out in the
following.

\begin{table}[t]
  \begin{center}
    \caption{%
      A comparison between \systemname{} and a selection of state-of-the-art
      fuzzing approaches proposed for \ac{MCU} firmware.
      Abbreviations: \textbf{E}mulated; \textbf{N}ative;
      \textbf{H}igh-level (function-level); \textbf{L}ow-level
      (\ac{MMIO}/register-level); \textbf{P}assthrough; \textbf{M}MIO;
      \textbf{D}MA; \textbf{H}ard\textbf{W}are; \textbf{S}ource \textbf{C}ode
    }
    \label{tab:compare}
    \small
    \begin{tabular}{ c|ccccccc|g| }
      Objectives &
      \rotatebox[origin=c]{90}{~HALucinator~\cite{clements2020halucinator}}    &
      \rotatebox[origin=c]{90}{~P2IM~\cite{feng2020p2im}}                      &
      \rotatebox[origin=c]{90}{~Para-rehosting~\cite{li2021library}}           &
      \rotatebox[origin=c]{90}{~Fuzzware~\cite{scharnowski2022fuzzware}}       &
      \rotatebox[origin=c]{90}{~$\mu$EMU~\cite{zhou2021automatic}}             &
      \rotatebox[origin=c]{90}{%
        Avatar\textsuperscript{2}~\cite{muench2018avatar}%
      } &
      \rotatebox[origin=c]{90}{Safirefuzz~\cite{seidel2023forming}}            &
      \rotatebox[origin=c]{90}{~\systemname{}}                         \\
      \midrule
      Code execution      & E    & E & N     & E & E & E & N    & N    \\
      Peripheral modeling & H    & L & H     & L & L & P & H    & H    \\
      Supported \ac{I/O}  & M, D & M & M, D  & M & M & M & M, D & M, D \\
      HW independence     & \cmark & \cmark & \cmark & \cmark &
                                     \cmark & \xmark & \cmark & \cmark \\
      SC independence     & \cmark & \cmark & \xmark & \cmark &
                                     \cmark & \cmark & \cmark & \cmark \\
      \bottomrule
    \end{tabular}
  \end{center}
\end{table}

\parasum{Hardware-in-the-loop is slow and doesn't scale $\Rightarrow$
software-only approach of EmbedFuzz solves both}
Hardware-in-the-loop solutions~\cite{chen2018iotfuzzer,corteggiani2018inception,
kammerstetter2016embedded,kammerstetter2014prospect,koscher2015surrogates,
zaddach2014avatar,talebi2018charm,muench2018avatar,gustafson2019analysis}
rely on the actual \ac{MCU} hardware.
Such solutions are inherently limited in performance due to communication
overhead with the \ac{MCU} through debug probes and provide poor scalability
due to additional hardware costs and configuration efforts.
\systemname{} overcomes those limitations thanks to peripheral modeling in
software and native code execution in a Linux user space process.

\parasum{Other software-modeling approaches lack accuracy, performance, bug
finding capabilities, or a combination of those}
Some re-hosting approaches propose modeling peripherals in software at the
kernel level, providing data through generic POSIX interfaces to a userspace
process~\cite{jiang2021ecmo,costin2016automated,kim2020firmae,zheng2019firmafl,chen2016automated}.
In contrast to \systemname{}, these approaches are not applicable to \ac{MCU}
firmware due to their focus on Linux-based embedded firmware.

Symbolically modeling values read from peripheral
registers~\cite{zhou2021automatic,hernandez2017firmusb,
shoshitaishvili2015firmalice,cao2020deviceagnostic,scharnowski2022fuzzware}
enables fully automated \ac{MCU} firmware re-hosting.
However, as stated by Fasano~\etal{}~\cite{fasano2021sok}, this technique often
over-approximates peripheral capabilities even for simple firmware binaries.
We argue that \systemname{}'s manual handler implementations provide higher
accuracy for possible values returned from peripherals and do not require to
run an analysis phase for each firmware to determine an appropriate peripheral
model.
Modeling peripheral registers by heuristically matching \ac{MMIO} access
patterns to common static register
types~\cite{feng2020p2im,harrison2020partemu,mera2021dice} requires running
costly analyses on each new firmware image similar to symbolic modeling
approaches.

Alternatively, other approaches~\cite{clements2020halucinator,clements2021your,
maier2020basesafe} model the peripheral behaviors at higher abstraction layers.
They leverage \ac{HAL} library functions for re-hosting and analyzing \ac{MCU}
firmware binaries by intercepting such high-level functions instead of
low-level \ac{MMIO} or register accesses.
These approaches leverage \ac{ISA} translation in an emulator to execute
target code, which incurs high runtime overhead and low fuzzing throughput.
In contrast, \systemname{} executes Cortex-M firmware on Cortex-A processors at
native speed using transplantation.

In concurrent work, Safirefuzz~\cite{seidel2023forming} (to be published in
USENIX Security '23) also leverages the similarity of the \arm{} Cortex-M and
Cortex-A \acp{ISA}, thereby achieving performance gains in fuzzing campaigns.
In contrast to \systemname{}, Safirefuzz fails to address semantic differences
in the \acp{ISA}, threatening the semantical correctness of executed code.
While Safirefuzz's dynamic binary rewriting approach at runtime is targeted
purely at fuzzing, \systemname{}'s combination of static binary rewriting and a
minimal runtime generically transplants firmware binaries into a Linux user
space process for any dynamic analysis use case.

Source code based approaches such as para-rehosting~\cite{li2021library} employ
manually implemented peripheral handlers directly compiled into the final
binary.
This approach relies on firmware source code, which is typically not provided
for \ac{COTS} \ac{IoT} devices.
In contrast, \systemname{} works on binary firmware without introducing
dependencies on the source code.

\section{Conclusion}
\label{sec:conclusion}

\systemname{} is the first end-to-end dynamic analysis framework
for \ac{MCU} firmware capable of running its targets at native speed.  We
implemented \systemname{} following our concept of \textit{firmware
transplantation}, a novel technique for re-hosting and a faster alternative to
previous emulation-based approaches.  Firmware transplantation exploits the
\ac{ISA} compatibility of low-end and high-end platforms such as \arm{} Cortex-M
and \arm{} Cortex-A, respectively. Our system adapts and extends
\acf{HLM} to allow for fuzzing the firmware without real hardware. For our
prototype, we employ \acs{AFL}++ as a drop-in replacement for a modern
general-purpose fuzzer.
Our evaluation results show that \systemname{} outperforms the state-of-the-art
in terms of performance by a factor of up to eight, resulting in efficiency
gains by a factor of at least four.
We highlight the importance of semantically correct execution of firmware code,
which \systemname{} provides through transplantation.
Finally, our bug triaging case studies show that \systemname{} facilitates the
analysis of bug root causes by enabling existing \ac{OS}-provided introspection
capabilities.

\bibliographystyle{ACM-Reference-Format}
\interlinepenalty=10000
\bibliography{main}


\appendix

\section{Instructions Rewritten by \systemname{}}
\label{app:semantically-different-instructions}

\systemname{} needs to handle a small set of instructions exhibiting different
behavior in an \arm Cortex-A user space process in comparison to an \arm
Cortex-M bare-metal firmware binary.
We categorize these instructions as follows:

\paragraph{Modifying \ac{MCU}-specific \ac{CPU} state.}
The instructions \texttt{msr} (move to special register from register) and
\texttt{mrs} (move to register from special register) copy information to and
from model-specific registers, respectively.
We rewrite those instructions to either only change arguments with semantic
equivalence (e.g., model the Cortex-M \acf{APSR} with Cortex-A's \acf{CPSR}) or
to emulate their behavior in instructions (e.g., copies to/from unused floating
point registers for modeling banked stack pointer registers).
Where direct replacement is not possible due to complex logic, we replace the
instruction with a \texttt{bkpt} instruction to trap-and-emulate in our
runtime.

\paragraph{Software interrupts.}
The \texttt{svc} instruction both on \arm{} Cortex-M and Cortex-A causes a
software interrupt to be raised.
Whereas the firmware expects the corresponding interrupt handler to be
executed, such a software interrupt in a Linux user space process causes a
transition to the kernel and execution of a syscall.
In order to properly execute the firmware's interrupt handler and not issue
unexpected syscalls on our fuzzing system, we replace \texttt{svc} instructions
with \texttt{bkpt} instructions and forward control flow to the interrupt
handler from our runtime's trap handler triggered as a result of the
\texttt{bkpt}.

\paragraph{Coprocessor accesses.}
On \arm{} \acp{MCU}, up to 16 coprocessors can be defined for adding features
to the main processor.
Any instruction executing on the main \ac{CPU} is also streamed to a
coprocessor and executed on the coprocessor if necessary~\cite{arm2021armv7m}.

Hardware floating point support implemented on a coprocessor does not require
special handling by \systemname{}, since the corresponding floating point
instructions are transparently executed on the floating-point implementation of
the host Cortex-A \ac{CPU}.

In our data set (see \cref{sec:evaluation}), we did not encounter custom
coprocessors.
In case a custom coprocessor is employed, the instructions \texttt{cdp/cdp2}
(coprocessor data processing), \texttt{mcr/mcr2} (move to coprocessor from
\arm{} register), \texttt{mcrr/mcrr2} (move to coprocessor from two \arm{}
registers), \texttt{mrc/mrc2} (move to \arm{} register from coprocessor),
\texttt{mrrc/mrrc2} (move to two \arm{} registers from coprocessor),
\texttt{ldc/ldc2} (load coprocessor), and \texttt{stc/stc2} (store coprocessor)
need to be rewritten accordingly.
Depending on the coprocessor's functionality, those instructions can either be
rewritten inline with equivalent functionality, or trap into the runtime for
corresponding emulation.

\end{document}